\documentstyle[epsfig,11pt]{article}
\newcommand{\be}{\begin{equation}}
\newcommand{\ee}{\end{equation}}
\newcommand{\ben}{\begin{eqnarray}}
\newcommand{\een}{\end{eqnarray}}
\newcommand{\bc}{\begin{center}}
\newcommand{\ec}{\end{center}}

\begin{document}

%\draft \twocolumn[\hsize\textwidth\columnwidth\hsize\csname
%@twocolumnfalse\endcsname \widetext

\title{Nearby quasar remnants and
ultra-high energy cosmic rays}
\author{Diego F. Torres$^1$, Elihu Boldt$^2$, Timothy Hamilton$^{2,3}$,
and Michael Loewenstein$^{2,4}$
%\address{
\\{\small $^1$Physics Department, Princeton University, NJ 08544, USA }\\
  {\small $^2$Laboratory for High Energy Astrophysics, NASA Goddard
              Space Flight Center, Greenbelt, MD 20771, USA}\\
  {\small $^3$National Research Council Associate}\\
  {\small $^4$Department of Astronomy, University of Maryland,
              College Park, MD 20742}
 }

\date{\today }

\maketitle

\begin{abstract}
  As
recently suggested, nearby quasar remnants are plausible sites of
black-hole based compact dynamos that could be capable of
accelerating ultra-high energy cosmic rays (UHECRs). In such a
model, UHECRs would originate at the nuclei of nearby dead
quasars, those in which the putative underlying supermassive black
holes are suitably spun-up.  Based on galactic optical luminosity,
morphological type, and redshift, we have compiled a small sample
of nearby objects selected to be highly luminous, bulge-dominated
galaxies, likely quasar remnants.  The sky coordinates of these
galaxies were then correlated with the arrival directions of
cosmic rays detected at energies $> 40$ EeV.  An apparently
significant correlation appears in our data. This correlation
appears at closer angular scales than those expected when taking
into account the deflection caused by typically assumed IGM or
galactic magnetic fields over a charged particle trajectory.
Possible scenarios producing this effect are discussed, as is the
astrophysics of the quasar remnant candidates. We suggest that
quasar remnants be also taken into account in the forthcoming
detailed search for correlations using data from the Auger
Observatory.\\

PACS number(s): 98.70.Sa, 98.54.-h

\end{abstract}

\section{Introduction}

Different experiments over the past few decades have detected several
giant air showers, confirming the arrival of cosmic rays (CRs) with
energies up to a few hundred EeV (1 EeV $\equiv 10^{18}$
eV)\cite{YD}.
%\footnote{Recent corrections in the primary energies of
%AGASA and Haverah Park events are discussed later.}
The nature and
origin of these energetic particles remain a mystery \cite{BS}. CRs,
however, can not travel unaffected through inter-galactic space. The
thermal photon background becomes highly blueshifted for
ultra-relativistic protons, and the reaction $p\gamma \rightarrow
\Delta^+ \rightarrow \pi^0 p$, and similar others, effectively degrade
the primary proton energy.  This provides a strong constraint on the
proximity of CR-sources, discovered early on by Greisen, and by
Zatsepin \& Kuz'min, and referred to as the GZK cutoff
\cite{gzk}. Specifically, fewer than 20\% of 300 EeV (100 EeV) protons
can survive a trip of 18 (60) Mpc. For nuclei and photons the
situation is, in general, more drastic
\cite{puget}.\\

    The possible solutions to this puzzle appear to fall into three
broad categories, viz: 1) There are many nearby sources (e.g.,
based on a substantial present-epoch population of galaxies
hosting core supermassive black holes). 2) There are only a few
nearby sources (e.g. \cite{ES}) and particles are isotropized by
strong deflections in Galactic and/or extragalactic magnetic
fields of micro-Gauss strength, close to existing upper limits
(e.g., see \cite{mf}, also \cite{isola}). 3) The particles somehow
travel relatively unhindered through the Cosmic Microwave
Background radiation field, either by virtue of being some exotic
new kind of weakly photon-interacting entity, or by violating the
Lorentz symmetry of special relativity (e.g. \cite{SPE}), or by an
as yet unknown other effect.\\

    The arrival directions of the primary particles could be a useful
source of information about the origin(s) of CRs.  However, due to
the small number of events, particularly at the highest energies,
correlation studies are to be considered preliminary.  For
instance, the first five events observed with $E > 80$ EeV did in
fact point toward high redshift radio-loud quasars, astrophysical
environments that could well accelerate CRs above the GZK energies
via shock mechanisms (see Farrar and Biermann, in Ref.
\cite{SPE}).  However, with the inclusion of subsequent data, this
association now seems to have disappeared \cite{Sigl:2000sn}. \\
%Nevertheless, using the full sample
%of CRs detected at the highest energies (more than 50 events for
%AGASA), some emerging correlations may yet be suggestive enough as
%to discriminate among proposed models.\\

    In this communication we explore whether UHECRs can be
in any way related to present-epoch supermassive black holes.  To
that end, we analyze what we can learn from the arrival directions
of cosmic rays concerning their possible correlation with nearby
massive dark objects, candidate QRs (quasar remnants), and their
underlying astrophysics.\footnote{The term ``quasar remnants" was
introduced by Chokshi and Turner \cite{CT} to describe the
present-epoch population of dead quasars harboring supermassive
black hole nuclei.}  The compact dynamo model has been proposed as
a natural mechanism for accelerating cosmic rays in such
environments \cite{EG}.  In this model UHECRs are produced in
nearby dead quasars harboring spinning supermassive black holes.
The required emf is generated by the black hole induced rotation
of externally supplied magnetic field lines threading the horizon.
The observed flux of CRs would apparently drain only a negligible
amount of energy from the black hole dynamo, and particles up to
at least 100 EeV are expected. It is then interesting to ask if we
are able to see any correlation between the sky position of these
QRs and those of the highest energy CRs, assuming different simple
configurations for the intervening magnetic field.

\section{Quasar remnant candidates}

The Nearby Optical Galaxy (NOG) catalog of Giuricin et al.
\cite{giu} is a complete magnitude-limited (corrected blue total
magnitude $B\leq14$), distance-limited (redshift $z \leq 0.02$)
sample of several thousand galaxies of latitude $\left| b \right|
> 20^\circ$, with their morphology $T$ also provided \cite{marp}.  We shall impose
very restrictive selection criteria, those necessary for obtaining
candidate objects providing the most favorable setting for a black
hole based compact dynamo model of UHECR production \cite{EG}. The
key goal of our strategy is to select {\it a priori} an NOG
subsample of galaxies that are likely to be quasar remnants.
Therefore, we are seeking optically bright galaxies whose
luminosities are bulge dominated (e.g., giant ellipticals).
Imposing a GZK-related horizon of $\sim 50$ Mpc assures that there
are no quasars in our sample.  We already know that the best
determined (and most massive) black hole nuclei tend to be
associated with bulge luminosities corresponding to $M_B \sim -21$
and brighter (for tables, see Kormendy's website at
http://chandra.as.utexas.edu/$\sim$kormendy/wwwbhtable-tech or the
second reference in \cite{2}).\\

Firstly, then, we shall impose a cutoff in redshift, requiring all
galaxies in our sample to be within $z \leq 0.01$, which for $H_0
= 75$ km/s/Mpc corresponds to 40 Mpc. Since the CRs detected are
mostly in the northern hemisphere (except those coming from SUGAR,
which are not used in the present analysis), we shall impose the
restriction that all QRs have equatorial latitudes north of $-10$
deg.  We further require the absolute blue magnitude to be
brighter than $M_B = -21$ (for $H_0=75$ km/s/Mpc) and an RC3 ({\it
Third Reference Catalog of Bright Galaxies}) morphological type
$T<-3$.  A more negative $T$ indicates greater bulge prominence.
In order to correlate the UHECR arrival directions with their
putative QR origins, we require that these candidate QRs not lie
within rich clusters (those having more than 50 members).  Some of
the most massive QRs could reside in rich clusters of galaxies
(e.g., all four QRs considered by Boldt and Loewenstein in Ref.
\cite{EG} are in rich clusters), but the magnetic field strength
in those clusters is presumably several micro-Gauss \cite{clust};
hence, UHECRs originating in such QRs would be extremely deflected
away from their sources. In addition, there are only a handful of
galaxies within large clusters in the Giuricin et al.'s sample
that survive all other constraints in order to be declared
plausible quasar remnant candidates.
% CRs originating from such QRs in field
%galaxies, or in galaxies belonging to small clusters, would have a
%better chance of pointing back to their origins.
Using these selection criteria, we obtain a sample of 12 candidate
QRs. These candidate QRs are listed
in Table 1, and their astrophysical properties are analyzed below. \\

The UHECR sample used is that obtained with AGASA above 40 EeV
\cite{aga}. There are 38 such events at $\left| b \right| >
20^\circ$; the angular precision of their arrival directions was
estimated as a circle of radius 1.6 degrees.  For energies $> 100$
EeV we also consider the 7 events observed at $\left| b \right|
>20^\circ$ compiled by Sigl et al. \cite{Sigl:2000sn}, 4 of which
were obtained with AGASA.\footnote{See Table 1 in
\cite{Sigl:2000sn} for this compilation and the associated error
estimates.} We note that the statistical test we shall report was
blind, i.e. we did not know beforehand if any of the QR galaxies
were coincident with high energy cosmic rays. The simulation
technique is the same as that used in
Refs. \cite{Sigl:2000sn,Romero:1999tk,Romero:1999mg}.  \\

\begin{table}
\label{TT0}
   \centering
\caption{Sample of QR candidates, columns are the B1950
   coordinates ($\alpha,\delta$), name of the galaxy, richness
   of the group to which the galaxy pertains (number of members),
   galactic longitude and latitude, mean redshift,
   corrected redshift  (against Hubble distortions, using
   models of the peculiar velocity fields
   Marinoni et al. \cite{mari}), morphological type code (RC3), and
   absolute magnitude (for $H_0$ =75). This table is based on data
   from the NOG catalog of Giuricin et al. provided to us by C.
   Marinoni \cite{giu}. }
\begin{tabular}{lllllllllll}
\hline $\alpha$ & $\delta$ & Name & \# & $l$ &$b$ & $\left< cz \right>$ &
$cz$  & $T$ & $M_B$ \\
  &  & & &  & &  [km/s]& [km/s] &
&  &\\ \hline
   108.938  &      85.808     &    NGC 2300  &       8 &127.708& 27.809&
2214&  2559  & -3.6   &   -21.17\\
   136.937  &      60.244     &    NGC 2768  &       5 &155.492& 40.563&
1469&  2063  & -3.1   &   -21.68\\
   161.296  &      12.846     &    NGC 3379  &      24 &233.490& 57.634&
 732&  1217  & -4.0   &   -21.07\\
   168.880  &      59.060     &    NGC 3610  &       5 &143.540& 54.462&
1816&  2467  & -3.9   &   -21.24\\
   168.926  &      58.274     &    NGC 3613  &       4 &144.338& 55.099&
2072&  2733  & -4.1   &   -21.32\\
   181.407  &      65.450     &    NGC 4125  &       4 &130.187& 51.341&
1536&  2109  & -4.6   &   -21.94\\
   182.431  &      13.484     &    NGC 4168  &       9 &267.668& 73.337&
2092&  2811  & -4.2   &   -21.04\\
   187.227  &      26.049     &    NGC 4494  &       7 &228.618& 85.316&
1160&  1882  & -4.5   &   -21.59\\
   188.871  &      74.469     &    NGC 4589  &       1 &124.234& 42.898&
2131&  2617  & -4.1   &   -21.38\\
   191.502  &     -5.5282     &    NGC 4697  &      30 &301.632& 57.064&
1168&  1476  & -4.1   &   -21.33\\
   206.895  &      60.438     &    NGC 5322  &       4 &110.279& 55.494&
1947&  2505  & -4.4   &   -21.89\\
   225.987  &      1.7986     &    NGC 5846  &      13 &000.427& 48.797&
1577&  1894  & -4.2   &   -21.27\\
\hline
\end{tabular}
\end{table}

\begin{figure}[t]
\vspace{-1.5cm}
\begin{center}
\includegraphics[width=8cm,height=11cm]{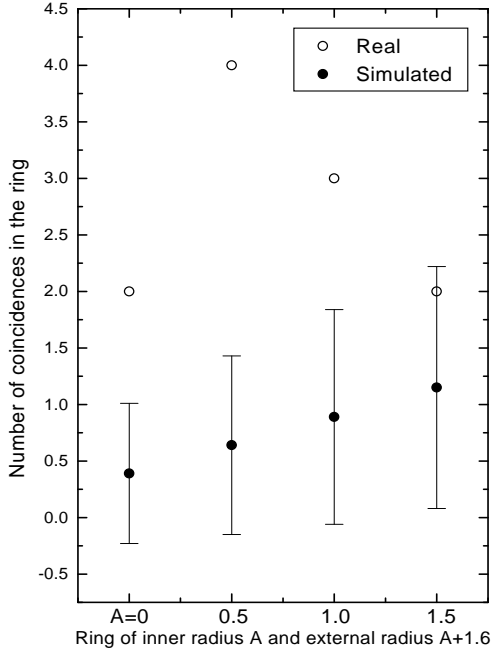}
\end{center}
\vspace{-1.5cm} \caption{Real coincidences in rings surrounding
the position of the central QR as a function of the internal
radius. In black we show the random result expectation, whereas in
light colour we show the real result. x-axis is given in degrees.
Error bars are 1$\sigma$.} \label{ring}
\end{figure}

%We first consider the case in which the uncertainty in the
%inferred CR source position on the sky arises solely from the
%experimental error in measuring the arrival direction.  Later, we
%will consider the case in which the uncertainty in the source
%position arises from a dispersion in arrival directions ({\it
%e.g.}, due to scattering in the intervening magnetic fields), as
%well as the measurement error.
Looking for superposed QRs in the cosmic ray error circles, we
find the results reported in Table \ref{P1}. When we consider the
AGASA sample with $E>40$ EeV, there is an excess in the real
result which approaches $3\sigma$.
%Indeed, if we enlarge the nominal error radius by half
%a degree and one degree, respectively, the probability for the
%real result to be an effect of chance diminishes.
This could argue in favor of a correlation between our sample of
QRs and UHECRs at small angular scales.  This is not the case for
the CRs with energies above 100 EeV, for which we found only an
apparent excess for the nominal error circles, not significant
enough to give to it any confidence. A summary of these results is
given in Figure \ref{ring}. There we show the number of real
coincidences in rings surrounding the position of the central QR
as a function of the internal radius of the ring, $A$. Each ring
has an angular width of 1.6 deg, and the number of coincidences is
shown with angular offsets between $A$ and $A+1.6$.  An excess of
coincidences is clearly evident for the innermost radii.  At
larger radii, the real superpositions are compatible with the
random coincidences. This result stands disregarding the value
chosen for the size
of the ring.\\

\begin{figure}[t]
\begin{center}
\includegraphics[width=8cm,height=11cm]{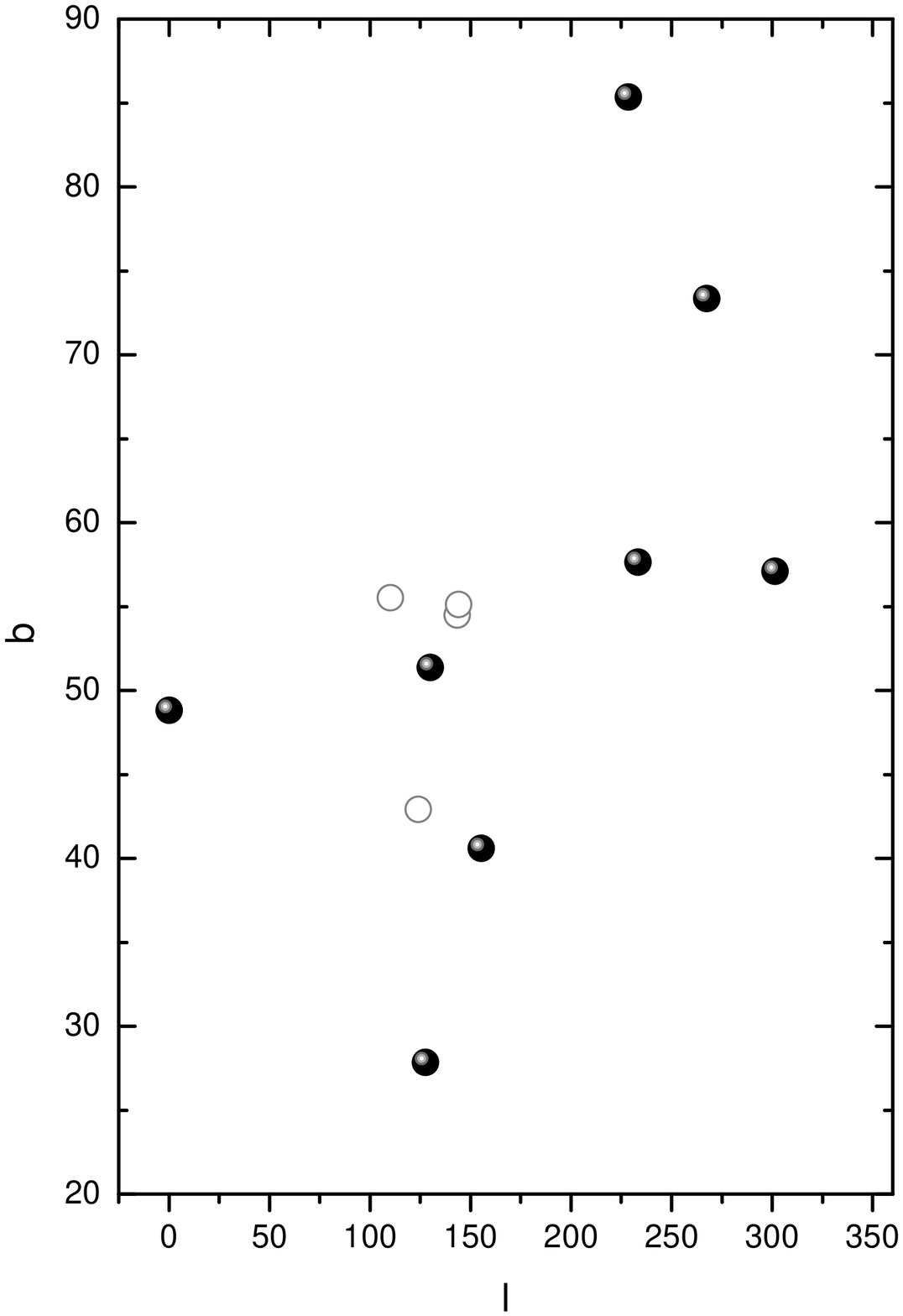}
\end{center}
\vspace{-1.5cm} \caption{Sky distribution (galactic coordinates)
of the selected QRs. The lighter dots stand for those QRs which
exhibit superposition with UHECRs. Two closely neighboring QRs (in
projection) are actually almost indistinguishable in this plots
(NGC 3610 and NGC 3613).} \label{1}
\end{figure}

Note that when increasing the error circles, no additional QRs are
involved in the coincidences, but instead, there are new CRs
involved.  This is what makes the real result increase.  Details
of the coinciding pairs are given in Table \ref{P2}.  {\it It is
risky, of course, to evaluate the level of significance that we
should attribute to this correlation, especially when so few
events/objects constitute the samples involved.}
 The 12 galaxies, however, are selected
{\it a priori} on a restrictive physical basis, aimed at
identifying QRs harboring supermassive black holes. Furthermore,
we can see that the correlated pairs identify an apparently
preferred region of the sky, as Figure \ref{1} shows. The expected
number of random coincidences decreases from 0.40 to 0.14 when the
region of the sky is restricted to an area given roughly by $160 <
\alpha < 200$ and $50 < \delta <80$.  This is $\sim 5 \sigma$
below the real result of 2 coincidences. This latter result,
however, is obtained {\it a posteriori}, and its importance is
thereby reduced. \\

\begin{table}
   \caption{Coincidences with the 34 AGASA CRs with energies of
   $E>40$ EeV and the given constraints in both $\delta$ and
   $b$. The 12 QRs given in Table 1 are considered.  The second panel
   shows similar results, but in this case, we consider the
   coincidences with 7 UHECR with $E>100$ EeV in the same latitude
   range. A real result equal to 2 means that there are 2 different
   cosmic rays coinciding with the quoted QRs.  }\label{P1} \centering
   \vspace{0.2cm}
\begin{tabular}{lclll}
\hline
Error considered & Real Result & Random Result & Poisson
Prob. & Galaxies involved\\\hline Nominal           & 2       &
0.40$\pm$0.60 & 0.04 (2.7$\sigma$)&
NGC 3610, 3613, 5322\\
Nominal + 0.5 deg & 3       & 0.68$\pm$0.82 & 0.02
(2.8$\sigma$)&NGC 3610, 3613, 5322\\
Nominal + 1.0 deg & 4       & 1.06$\pm$1.03 & 0.01
(2.9$\sigma$)&NGC 3610, 3613, 5322\\
\hline \hline
Nominal           & 1       & 0.23$\pm$0.45 & 0.17 (1.7$\sigma$)& NGC 4589\\
Nominal + 0.5 deg & 1       & 0.30$\pm$0.52 & 0.22 (1.1$\sigma$)& NGC 4589\\
Nominal + 1.0 deg & 1       & 0.40$\pm$0.61 & 0.26 (1.0$\sigma$)& NGC 4589\\
\hline
\end{tabular}
\end{table}

\begin{table}
   \centering
   \caption{Superposed pairs of QRs and CRs. Nominal errors are
   considered. $\theta(E)$ is given for the nominal values in its
   definition.  The offset angles for NGC 3610 and NGC 3613 refer to the
   same CR, which lies within both error circles.
    }
   \vspace{0.2cm}
   \label{P2}
   \begin{tabular}{lllll}
\hline Galaxy & CR Energy & Experiment & $\theta(E)$ & Angular\\
& $10^{19}$ eV  &  & &offset\\\hline
NGC 3610 & 7.7 & AGASA & 3.7 & 1.3 \\
NGC 3613 & 7.7 & AGASA & 3.9 & 0.7 \\
NGC 5322 & 4.4 & AGASA & 6.6 & 0.8 \\\hline
\end{tabular}
\end{table}

If we consider that charged particles (protons) are accelerated
in these QRs and then travel through the intergalactic and Galactic
magnetic fields towards the Earth, we expect their trajectories to be
deflected.  When the Larmor radius of a particle ($r_{\rm L} \simeq
10^2$ Mpc $E_{20}/B_{-9}$) is much larger than the coherence length of
the magnetic field $\ell_{\rm coh}$, the characteristic deflection
angle $\theta$ from the direction of the source, located at a distance
$D$, can be estimated assuming that the particle makes a random walk
in the magnetic field (see e.g. \cite{we})
\begin{equation}
   \theta(E) \simeq 3.8^\circ\, \left(\frac{D}{50\,\,{\rm
   Mpc}}\right)^{1/2}\,\, \left(\frac{\ell_{\rm coh}}{1 \,\, {\rm
   Mpc}} \right)^{1/2} \,\, \left(\frac{B_{-9}}{E_{20}} \right)\,\,,
\end{equation}
where $E_{20}$ is the energy of the particle in units of
$10^{20}$eV, and $B_{-9}$ is the magnetic field in units of
$10^{-9}$ G. We can see that scattering in large scale magnetic
irregularities ${\cal O}$ (nG) \cite{b} are enough to bend the
orbits of trans-GZK protons by about $4$ deg in a 50 Mpc
traversal.  As exhibited in Table 3, the CR angular offsets
observed for these QRs are much smaller than $\theta$. For a
variety of assumed magnetic field scenarios, $\theta$ is often
substantially larger than the estimated AGASA measurement error of
1.6 degrees.  If, in fact, due to the value of the inter-galactic
magnetic field, such is the case, it would then indicate that
\begin{enumerate}
    \item the number of apparent QR/UHECR associations is no
    longer above random expectations (see Figure 1 for large
    value of $A$), and
    \item the {\it a priori} probability is relatively very
    small for offset angles as little as those actually observed
    (Table 3).
\end{enumerate}

Should the apparent clustering of correlated pairs be supported by
future data, what are the viable scenarios under which this could
occur? One possibility is to consider that the intergalactic
medium between Earth and the three apparently `clustered' QRs is
sufficiently different from the intergalactic medium in front of
the remaining nine objects that are much more uniformly
distributed on the accessible sky.  For the deflection of an
energetic (60 EeV, the mean of the two CR energies in Table 3)
proton in traversing 34 Mpc (the mean of the 3 QR distances in
Table 3) to be less than a degree, we would need $B <
2\times10^{-10}\;\ell_{\rm coh}^{-0.5}$ G, which appears to be not
as drastic a difference from the canonical nano-Gauss $B$ field
and coherence length of 1 Mpc that are usually assumed. Indeed,
the IGM B field is likely to be an order of magnitude less than a
nano-Gauss in voids comparable in size to the GZK horizon (Peter
Biermann; 2002, personal communication). Independent estimates of
the Intergalactic Medium (IGM) magnetic field towards these
galaxies would prove very useful in assessing this
explanation. \\

It is important to note that, in some directions, the magnetic
field of our own galaxy could well lead to a deflection of up to
several degrees for primaries with energies below 60 EeV; see, for
instance, Table 1 of Ref. \cite{stanev}. A recent study of this
issue was presented by Alvarez-Mu\~niz et al. \cite{m}. However,
it is important to realize that the possible filamentary topology
of the Galaxy's magnetic field (Gerrit Verschuur; 2002, personal
communication) would likely allow some directional windows, albeit
narrow, where the deflection of an UHECR could be much less than
typical. Deflections due to the magnetic fields would of course be
avoided if the primary were a photon, generated in the
neighborhood of the QR via an accelerated charged particle
interaction. A photon primary is ruled out for the highest cosmic
ray energy event detected with Fly's Eye \cite{ES}, but there is
yet some discussion of the likelihood of having some photons as
primaries in the AGASA sample \cite{g}. Protheroe and Johnson (see
second reference in \cite{puget}) have studied the propagation of
ultra-high energy gamma rays in the intergalactic magnetic field
in detail.  They have shown that if the intergalactic magnetic
field is low enough, say $B\sim 10^{-10}$ G, a photon can survive
a distance comparable to those of the QRs in our sample and be a
plausible primary for the observed showers. A new neutral hadron
could obviously be another extreme possibility. However, as we
have seen, the involved QRs are all within a relatively small
region of the sky; their angular separation is clearly less than
the average angular separation between neighbors in the remaining
sample of nine.  Why would sources of neutral hadronic CRs be
restricted to a limited angular region of the sky? Albeit
suggestive, we warn the reader that the effect observed here might
be no more than an artifact resulting from the small number
statistics available at this time. It is, however, interesting to
study the underlying astrophysics of our candidates as possible
cosmic rays emitters.

\section{Underlying Astrophysics}

Table 4 exhibits the physical attributes of our sample of QR
candidates, viz: morphological type, mass of supermassive black
hole nucleus ($M_{\rm BH}$), galactic bulge mass ($M_{\rm GB}$),
black hole accretion rate $\dot{m}$ (in Eddington units), the
maximum resulting proton energy ($E_{\rm max}$) expected, and
lower limits to the proton's radiation length ($\Lambda$) for
losses arising from photo-pion production in ultra-relativistic
collisions with the ambient electromagnetic radiation (photon)
field.  That the nuclei of these galaxies are indeed dark as well
as massive may be appreciated by noting that the radio luminosity
($\nu L_\nu$ at 6 cm) within a 5 arcsec beam \cite{nagar} is in no
instance greater than $10^{-8}$ (in Eddington units); the ROSAT
obtained X-ray luminosity \cite{1}, within a broader beam that
extends beyond the nucleus, is in every case already less than
$10^{-5}$ (in Eddington units).  The four QRs that appear to be
correlated with UHECRs ($ > 40$ EeV) are identified here by their
offset angles relative to the measured arrival direction of the
primary particle initiating the associated air shower; this offset
is never greater than the estimated experimental error in
determining this arrival direction. For the worst case, NGC 4589,
the associated Haverah Park UHECR
direction measured has an angular uncertainty radius of 5.6 degrees.\\

Along with the six published values cited in Table 4, we estimate
the black hole mass ($M_{BH}$) for all twelve QRs from the
velocity dispersion of stars \cite{2} within the bulge. The
galactic bulge mass ($M_{\rm GB}$) is derived from its stellar
luminosity using the mass--luminosity formula $\left( M/M_\odot
\right) / \left( L/10^{10} L_\odot \right)^{0.15}$ \cite{3}.  For
the S0 galaxy NGC 2300, we assume a typical S0 bulge--to--disk
ratio of $m_V{\rm (bulge)} - m_V{\rm (total}) = 0.60$ \cite{sdv}.
We note that the bulge mass loss rate in giant elliptical galaxies
is $\sim [M_{{\rm GB}} / (10^{12} M_\odot)] \, M_\odot / {\rm
year}$ \cite{4} and assume that the rate of accretion onto the
central black hole is an order of magnitude less, as typically
found for the Bondi rate \cite{5}; based on this, we estimate that
the accretion rate is ${\rm d}M / {\rm d}t \approx 0.1 M_{12} \,
(M_\odot / {\rm year})$, where $M_{12} \equiv [(M_{\rm GB}) /
(10^{12} M_\odot)]$.  In Eddington units, this rate is then
\begin{equation}
  \dot{m} \equiv c^2 ({\rm d}M / {\rm d}t) / L_{{\rm Edd}}
  \approx 0.45 (M_{12}/M_8) {\rm  ,}
\end{equation}
where $M_8 \equiv [(M_{{\rm BH}}) / (10^8 M_\odot)]$, and $L_{{\rm
Edd}} = 1.3 \times 10^{46} M_8$ ergs/s is the Eddington luminosity
limit.\\

From equations 2---4 in the last paper of Ref. \cite{EG}, we
obtain that, for losses dominated by curvature radiation, the
maximum proton energy expected via the dynamo action considered is
\begin{equation}
E_{\rm max} = 77 \, ({\rm d}M / {\rm d}t)^{1/8} \, \left(
M_8 \right)^{1/4} {\rm EeV} \, \approx \, 58 \, \left( M_{12}
\right)^{1/8} \, \left( M_8 \right)^{1/4} {\rm EeV.}
\end{equation}
We note that the values of $E_{\rm max}$ for all 12 galaxies in
Table~4 lie above 40 EeV, the lower limit characterizing the AGASA
sample considered.\\

A lower limit to the radiation length ($\Lambda_{\rm min}$) for
proton energy loss associated with photo-pion production is
estimated by considering the population of target photons within
the source region $\left[ R {\rm (source \; radius)} \geq 2GM/c^2
\right]$ at radio frequencies $\nu \geq 360(\gamma/10^{11})^{-1}$
GHz \cite{EG}, where $\gamma$ is the Lorentz factor given by
$\gamma = (E_{\rm max})/(938 {\rm  MeV})$.  The radio estimates
required are extrapolated from data at lower frequencies \cite{6}.
Those QRs with $\Lambda_{\rm min} > R$ are expected to
successfully accelerate protons up to $\sim E_{\rm max}$.\\

Apart from their apparent correlation with UHECRs, there is no
obvious systematic difference between the first four QR candidates
listed at the top of Table 4 and the eight remaining ones at the
bottom. However, it's important to note that we do not as yet know
the spin states of the supermassive black holes associated with
the nuclei of these galaxies; this key parameter might differ
substantially among them.  In each instance, the present state of
spin depends on the specific prior history involved (e.g.,
accretion evolution, merger interactions, and earlier activity).
Since the galactic nuclei considered here are X-ray dark, as are
most (see first Ref. in \cite{5}), they are not viable candidates
for black hole spin determination by means of the accretion disk
iron K-line x-ray florescence that appears to be so promising a
spectroscopic tool for active galactic nuclei, particularly
Seyferts \cite{8}.  We emphasize that a black hole state of near
maximal spin is a necessary condition for the realization of a
compact black hole dynamo of the sort envisaged for accelerating
UHECRs \cite{EG}. In this sense, if confirmed, the correlation of
UHECRs and QRs might well signal the introduction of a new means
for identifying those nearby isolated non-active galactic nuclei
that harbor highly spun-up black holes \cite{9}.

\begin{table}
\label{TTlast}
   \centering
   \caption{Physical attributes of quasar remnant candidates.
Col. (1), catalog number.  Col. (2), morphological type.  Col.  (3),
distance, $D$, corresponding to the corrected redshift in Table 1 (for
$H_0 = 75$).  Col. (4), galactic bulge mass, $M_{{\rm GB}}$, in units
of $10^{12} M_\odot$.  Col. (5), black hole mass, $M_{{\rm BH}}$, in
units of $10^8 M_\odot$.  Col. (6), accretion rate, $\dot{m}$, in
Eddington units.  Col. (7), maximum proton energy, $E_{{\rm max}}$.
Col. (8), minimum radiation length, $\Lambda_{{\rm min}}$, relative
to source size, $R$.  Col. (9), offset angle between the putative QR
and the most nearly aligned UHECR observed ($ > 40$ EeV).
} \vspace{0.2cm}

   \begin{tabular}{ccccccccc}
\hline NGC & Type & $D$ & $M_{\rm GB}$ & $M_{\rm BH}$ & $\dot m$ & $E_{\rm max}$ & $\Lambda_{\rm min}/R$ & UHECR Offset \\
     &    & Mpc& $M_{12}$ &$M_8$       &      & EeV   &       &degrees\\ \hline
3610 & E5 & 33 & 0.43 & 0.51           & 0.39 & 44    & 2.9   & 1.3  \\
3613 & E6 & 36 & 0.47 & 1.56           & 0.14 & 59    & 5.8   & 0.7  \\
4589 & E2 & 35 & 0.50 & 2.57  (4.71$^\dag$)  & 0.05 & 67 (78)  & 0.23 (0.38) & 4.7 \\
5322 & E3 & 33 & 0.85 & 2.63  (12.4$^\dag$)  & 0.03 & 72 (107) & 0.36 (1.3) & 0.8 \\

2300 & S0 & 34 & 0.19 & 4.28           & 0.02 & 68    & 14    &      \\
2768 & E6 & 28 & 0.68 & 1.56           & 0.20 & 62    & 0.5   &      \\
3379 & E1 & 16 & 0.36 & 1.71  (1.53$^\ddag$) & 0.11 & 58 (57) & 23 (21)  & \\
4125 & E6 & 28 & 0.90 & 2.79           & 0.15 & 74    & 15    &      \\
4168 & E2 & 37 & 0.35 & 0.984 (7.29$^\dag$)  & 0.02 & 51 (84) & 0.45 (2.3) & \\
4494 & E1 & 25 & 0.62 & 0.484 (8.58$^\dag$)  & 0.03 & 46 (94) & 4.5 (49)  &  \\
4697 & E6 & 20 & 0.47 & 0.866 (2.9$^\ddag$)  & 0.07 & 51 (69) & 10 (27)  &   \\
5846 & E0 & 25 & 0.44 & 4.1            & 0.05 & 75    & 2.6   &      \\
\hline
\end{tabular}

\vspace{0.2cm}

\begin{flushleft}Note---Black hole masses in the table are normally
calculated from stellar velocity dispersions.  Those black hole
masses taken from the literature are shown in parentheses, as are
the quantities derived from them.  A $\dag$ symbol stands for
masses taken from van der Marel 1999, whereas a $\ddag$ symbol,
for masses taken from Kormendy \& Gebhardt 2001 \cite{bhmass}.
These literature masses are corrected for our assumed distances.
\end{flushleft}

\end{table}

\section{Concluding remarks}
As recently suggested, quasar remnants are plausible sites of
black-hole based compact dynamos that could be capable of
accelerating protons up to ultra-relativistic energies. We have
found that nearby quasar remnant candidates present an
above-random positional correlation with the sample of UHECRs. The
correlation appears on closer angular scales than those expected
when taking into account the deflection caused by typically
assumed intergalactic or Galactic magnetic fields. Possible
scenarios producing this effect were discussed; if real, the
plausible fine structure of the Galactic field may ultimately
provide the basis for the most natural explanation. In order to
substantiate and further investigate the apparent correlation
reported here between QR candidates and CR arrival directions, we
need a large, reliable sample of the most energetic UHECRs, those
of the very highest magnetic rigidity. It is hard to claim a
definitive correlation with few objects and CRs constituting the
samples. Future experiments, such as the Pierre Auger observatory
\cite{cronin}, EUSO \cite{euso} and the NASA space-borne OWL
(Orbiting Wide-angle Light-collectors) mission \cite{streit},
should vastly increase the availability of UHECRs $\geq 100$ EeV.
As we have already noted, however, CR sources in rich galactic
clusters (those with pervasive micro-Gauss fields) are not well
suited for our present kind of correlative investigation, even
though the nearest QR candidates of interest reside in such
systems. We suggest that QR candidates located within the
relatively nearby Virgo and Fornax clusters \cite{EG} might be
best studied by means of the TeV curvature radiation expected from
the putative compact black-hole dynamos associated with these
objects (see second paper in \cite{EG}). This work suggests that
QRs should also be taken into account when analyzing coincidences
in the forthcoming Auger Observatory.

\subsection*{Acknowledgments}

DFT acknowledges L. Anchordoqui and G. Sigl for discussions and
criticism, and support from Fundaci\'on Antorchas, CONICET, and
Princeton University. Much appreciated questions posed by the
referee have led to substantial improvement of our paper.


\begin{thebibliography}{}


\bibitem{YD} S. Yoshida, and H. Dai, J. Phys. G {\bf 24}, 905
(1998); M.~Nagano and A.~A.~Watson,
%``Observations And Implications Of The Ultrahigh-Energy Cosmic Rays,''
Rev.\ Mod.\ Phys.\ {\bf 72} (2000) 689. %%CITATION = RMPHA,72,689;%%


\bibitem{BS} P. Bhattacharjee, and G. Sigl, Phys. Rep. {\bf 327},
109 (2000), and references therein.


\bibitem{gzk} K. Greisen, Phys. Rev. Lett. {\bf 16}, 748 (1966); G.
T. Zatsepin, and V. A. Kuz'min, Pis'ma Zh. \'Eksp. Teor. Fiz. {\bf
4}, 114 (1966) [JETP Lett. {\bf 4}, 78 (1966)].


\bibitem{puget} J. L. Puget, F. W. Stecker and J. H. Bredekamp,
ApJ {\bf 205}, 638 (1976); R. J. Protheroe and P. Johnson,
Astropart. Phys. {\bf 4}, 253 (1996); F.~W.~Stecker and
M.~H.~Salamon,
%``Photodisintegration of ultrahigh energy cosmic rays: A new
%determination,''
ApJ {\bf 512}, 521 (1992)
[arXiv:astro-ph/9808110]. %%CITATION = ASTRO-PH 9808110;%%


\bibitem{ES}J. W. Elbert and P. Sommers, ApJ
{\bf 441}, 151 (1995).


\bibitem{mf}J. Wdowczyk, and A. W. Wolfendale, Nature {\bf 281}, 356
(1979); G. Sigl, M. Lemoine, and P. Biermann, Astropart. Phys.
{\bf 10}, 141 (1999); M. Lemoine, G. Sigl, and P. Biermann,
[arXiv:astro-ph/9903124]; G.~R.~Farrar and T.~Piran,
Phys.~Rev.~Lett. {\bf 84}, 3527 (2000); P.~Blasi, S.~Burles, and
A.~V.~Olinto, ApJ {\bf 514}, L79 (1999);L. A. Anchordoqui, H.
Goldberg and T. J. Weiler, Phys. Rev. Lett. {\bf 87}, 081101
(2001) [arXiv:astro-ph/0103043]; L. Anchordoqui, H. Goldberg, S.
Reucroft and J. Swain, [hep-ph/0107287].

\bibitem{isola} C. Isola, and G. Sigl [arXiv:astro-ph/0203273].

\bibitem{SPE} G. R. Farrar, and P. L. Biermann, Phys. Rev.
Lett. {\bf 81}, 3579 (1998); A. Virmani, S. Bhattacharya, P. Jain,
S. Razzaque, J. P. Ralston, and D. W. McKay,
[arXiv:astro-ph/0010235]; P. G. Tinyakov, and I. I. Tkachev,
[arXiv:astro-ph/0102476]; L. A. Anchordoqui et al., Mod. Phys.
Lett. A{\bf 16}, 2033 (2001); T. J. Weiler, Astropart. Phys. {\bf
11}, 303 (1999); {\it ibid} {\bf 12}, 379E (2000); D. Fargion, B.
Mele and A. Salis, ApJ {\bf 517}, 725 (1999); Gonzales-Mestres, in
Proceedings of the 25th ICRC, Durban, So. Africa, 1997, edited by
M.S. Polgieter, B.C. Raubenheimer, and D.J. van der Walt (World
Scientific, Singapore, 1997); S. Coleman and S.L. Glashow, Phys.
Rev. D59, 116008 (1999); C.~Tyler, A.~V.~Olinto and G.~Sigl,
%``Cosmic neutrinos and new physics beyond the electroweak scale,''
Phys.\ Rev.\ D {\bf 63}, 055001 (2001) [hep-ph/0002257];
%%CITATION = HEP-PH 0002257;%%
L.~Anchordoqui, H.~Goldberg, T.~McCauley, T.~Paul, S.~Reucroft and
J.~Swain,
%``Extensive air showers with TeV-scale quantum gravity,''
Phys.\ Rev.\ D {\bf 63}, 124009 (2001) [hep-ph/0011097].
%%CITATION = HEP-PH 0011097;%%


%\cite{Sigl:2000sn}
\bibitem{Sigl:2000sn}
G.~Sigl, D.~F.~Torres, L.~A.~Anchordoqui and G.~E.~Romero,
%``Testing the correlation of ultra-high energy cosmic rays with high  redshift sources,''
Phys.\ Rev.\ D {\bf 63}, 081302 (2001) [arXiv:astro-ph/0008363].
%%CITATION = ASTRO-PH 0008363;%%


\bibitem{EG} E. Boldt, and P. Ghosh,
Mon.\ Not.\ Roy.\ Astron.\ Soc.\ {\bf 307}, 491 (1999); A.
Levinson Phys. Rev. Lett. {\bf 85}, 912 (2000); E. Boldt and M.
Loewenstein, Mon.\ Not.\ Roy.\ Astron.\ Soc.\ {\bf 316}, L29
(2000).


\bibitem{CT} A. Chokshi and E. L. Turner,  Mon.\ Not.\ Roy.\ Astron.\ Soc.\ {\bf 259}, 421, 1992.


\bibitem{giu} G. Giuricin, C. Marinori, L. Ceriani, and A. Pisani,  ApJ
{\bf 543}, 178 (2000).


\bibitem{clust} T. E. Clarke, P. P. Kronberg, ApJ {\bf 547}, L111
(2001).


\bibitem{marp} C. Marinoni, private communication.


\bibitem{mari} C. Marinoni et al. ApJ {\bf 501}, 484 (1998).


\bibitem{we} E. Waxman and J. Miralda-Escud\`e, ApJ
{\bf 472}, L89 (1996).


\bibitem{b} See, for instance, L. A. Anchordoqui
and H. Goldberg, [hep-ph/0106217] and references in \cite{mf}
above.


\bibitem{aga} M. Takeda et al., ApJ {\bf 522}, 225 (1999)
[arXiv:astro-ph/9902239]; [arXiv:astro-ph/0008102].


%%\bibitem{HP} M. Ave, J. Knapp, J. Lloyd-Evans, M. Marchesini, and A.A.
%%Watson, astro-ph/0112253.


\bibitem{AGA}N. Sasaki et al., Proceedings of the ICRC 2001, in
press.


%\cite{Romero:1999tk}
\bibitem{Romero:1999tk}
G.~E.~Romero, P.~Benaglia and D.~F.~Torres,
%``Unidentified 3EG gamma-ray sources at low galactic latitude,''
Astron.\ Astrophys.\  {\bf 348}, 868 (1999)
[arXiv:astro-ph/9904355].
%%CITATION = ASTRO-PH 9904355;%%

%\cite{Romero:1999mg}
\bibitem{Romero:1999mg}
G.~E.~Romero, D.~F.~Torres, I.~Andruchow, L.~A.~Anchordoqui and
B.~Link,
%``Gamma Ray Bursts with peculiar temporal asymmetry,''
Mon.\ Not.\ Roy.\ Astron.\ Soc.\  {\bf 308}, 799 (1999)
[arXiv:astro-ph/9904107].
%%CITATION = ASTRO-PH 9904107;%%



\bibitem{stanev} T. Stanev, ApJ {\bf 479}, 290 (1997).


\bibitem{m} J. Alvarez-Mu\~niz, R. Engel, and T. Stanev,
[arXiv:astro-ph/0112227].


\bibitem{g} K. Shinozaki et al., Proceedings of the ICRC 2001, in
press.


\bibitem{nagar} Nagar, N. M., Wilson, A.
S. \& Falcke, H., ApJ {\bf 559}, L87 (2001).


\bibitem{1}E. O'Sullivan, D. A. Forbes, \& T. J Ponman,
[arXiv:astro-ph/0108181].


\bibitem{2}D. B. McElroy, ApJS {\bf 100}, 105 (1995); D. Merritt,
\& L. Ferrarese, ApJ {\bf 547}, 140 (2001).


\bibitem{3} S. M. Faber, et al., AJ {\bf 114}, 1771 (1997); A. Wandel,
ApJ {\bf 519}, L39 (1999).


\bibitem{sdv}F. Simien \& G. de Vaucouleurs, ApJ {\bf 302}, 564 (1986).


\bibitem{4}A. Athey, et al. [arXiv:astro-ph/0201338].


\bibitem{5}M. Loewenstein, et al. ApJ {\bf 555}, L21 (2001);
T. Di Matteo, et al. [arXiv:astro-ph/0202238].


\bibitem{6}M. Birkinshaw, \& R. L. Davies, ApJ {\bf 291}, 32 (1985);
T. Di Matteo, C. L. Carilli,  \& A. C. Fabian, ApJ {\bf 547}, 731 (2001);
P. B. Eskridge, G. Fabbiano, \& D-W. Kim, ApJS {\bf 97}, 141 (1995);
E. Hummel, J. M. van der Hulst, \& J. M. Dickey, A\&A {\bf 134}, 207
(1984); J. M. Wrobel, AJ {\bf 101}, 127 (1991); J. M. Wrobel, \& D. S.
Heeschen, AJ {\bf 101}, 148 (1991).


\bibitem{8}K.  Nandra, et al. ApJ {\bf 477}, 602 (1997).


\bibitem{9}E. Boldt, A. Levinson, and M. Loewenstein, Classical and
Quantum Gravity, 19, 1317 (2002).


\bibitem{bhmass}R. van der Marel, AJ {\bf 117}, 744 (1999); J. Kormendy
\& K. Gebhardt (2001) [arXiv:astro-ph/0105230].


\bibitem{cronin}J. Cronin, Rev. Mod. Phys. {\bf 71}, S165 (1999); J.
Cronin, in Unsolved Problems in Astrophysics, edited by J.N. Bahcall
\& J.P. Ostriker (Princeton Univ. Press, Princeton, 1997), p. 325.

\bibitem{euso} EUSO web pages: http://ifcai.pa.cnr.it/euso.html

\bibitem{streit}R. Streitmatter, in Workshop on Observing the Highest
Energy Particles From Space, AIP Proc. 433, edited by J. Krizmanic,
J. Ormes, \& R. Streitmatter (Am. Inst. Phys., New York, 1998), p. 95.

\end{thebibliography}
\end{document}